# A diamond AGPM coronagraph for VISIR

C. Delacroix*[a], O. Absil[b], D. Mawet[c], C. Hanot[b], M. Karlsson[d],
P. Forsberg[d], E. Pantin[e], J. Surdej[b] and S. Habraken[a]

[a]HOLOLAB, Université de Liège, 17,B5a Allée du 6 Aout, B-4000 Liège, Belgium
[b]IAGL, Université de Liège, 17,B5c Allée du 6 Aout, B-4000 Liège, Belgium
[c]ESO, Alonso de Córdova 3107, Vitacura, 763 0355 Santiago, Chile
[d]Ångström Laboratory, Uppsala University, Lägerhyddsvägen 1, SE-751 21 Uppsala, Sweden
[e]CEA Saclay, Orme des Merisiers Bat. 709, 91191 Gif sur Yvette Cedex, France

## ABSTRACT

In recent years, phase mask coronagraphy has become increasingly efficient in imaging the close environment of stars, enabling the search for exoplanets and circumstellar disks. Coronagraphs are ideally suited instruments, characterized by high dynamic range imaging capabilities, while preserving a small inner working angle. The AGPM (Annular Groove Phase Mask, Mawet et al. 2005[1]) consists of a vector vortex induced by a rotationally symmetric subwavelength grating. This technique constitutes an almost unique solution to the achromatization at longer wavelengths (mid-infrared). For this reason, we have specially conceived a mid-infrared AGPM coronagraph for the forthcoming upgrade of VISIR, the mid-IR imager and spectrograph on the VLT at ESO (Paranal), in collaboration with members of the VISIR consortium. The implementation phase of the VISIR Upgrade Project is foreseen for May-August 2012, and the AGPM installed will cover the 11-13.2 μm spectral range. In this paper, we present the entire fabrication process of our AGPM imprinted on a diamond substrate. Diamond is an ideal material for mid-infrared wavelengths owing to its high transparency, small dispersion, extremely low thermal expansion and outstanding mechanical and chemical properties. The design process has been performed with an algorithm based on the rigorous coupled wave analysis (RCWA), and the micro-fabrication has been carried out using nano-imprint lithography and reactive ion etching. A precise grating profile metrology has also been conducted using cleaving techniques. Finally, we show the deposit of fiducials (i.e. centering marks) with Aerosol Jet Printing (AJP). We conclude with the ultimate coronagraph expected performances.

## 1. INTRODUCTION

Direct imaging of a very high contrast scene requires to cancel the bright source, or at least drastically reduce its intensity, in order to characterize the faint sources situated in the close environment of the bright astronomical object. Especially for extrasolar planets, very high rejection ratios are necessary, because the faint companion is $10^4$-$10^7$ times fainter than its parent star in the thermal infrared (L-band, M-band, N-band, Q-band, from 3.5 to 20 μm). Therefore, such direct detection of exoplanets necessitates dedicated instruments such as coronagraphs. The first coronagraph has been proposed by Bernard Lyot (Lyot 1939[2]) and consists of an opaque spot centered on the optical axis. This Lyot coronagraph, considered as an amplitude coronagraph, was dedicated to observe the corona of the sun, but when applied to stellar observation, it occults necessarily an interesting zone around the star (several $\lambda/D$) including the potential exoplanets behind it. Since then, amplitude based coronagraph techniques have evolved allowing an improved contrast performance (Soummer et al. 2003[3], Kuchner et al. 2002[4]). Several extrasolar planets have been directly imaged (Fomalhaut, HR 8799, Beta Pictorisb, Kalas et al. 2008[5], Marois et al. 2008[6], Lagrange et al. 2009[7]). However, these observations were made under exceptional circumstances: especially large planets (considerably larger than Jupiter), generally warm, and widely separated from their host star. Indeed, this family of coronagraphs is still strongly limited in throughput and inner working angle.

*cdelacroix@ulg.ac.be ; www.hololab.ulg.ac.be

To overcome these limitations and enable the observation of smaller exoplanets (i.e. earth-like), an alternative approach to amplitude coronagraphs consists of varying the phase of the incident light instead of blocking it. Such a device is called a phase mask coronagraph (Roddier & Roddier 1997[8], Rouan et al. 2000[9]). A destructive interference can be achieved by combining two portions of the incident light beam, one of which having to undergo a half-wavelength (or π) phase shift, which causes the nulling of the bright source in the relayed pupil downstream of the coronagraph. The very last evolution of this phase mask is the vector vortex coronagraph (VVC), i.e. a phase mask in which the phase shift varies azimuthally around the center. This technique has been experimentally demonstrated using liquid crystal polymer (Mawet et al. 2010[10], Serabyn et al. 2010[11]) or photonic crystals (Murakami et al. 2010[12]). A weakness of these liquid or photonic crystal coronagraphs is their bandwidth: they do not transmit at wavelengths beyond the near-infrared (H-band centered at ~1.65 μm, K-band ~2.2 μm). Meanwhile, the demand for efficient coronagraphs in the mid-infrared is increasing, following the recent success of high contrast imaging of exoplanets and circumstellar disks in the L-band, N-band and Q-band (e.g. Lagrange et al. 2010[13], Moerchen et al. 2007[14]). Therefore, we are pursuing a different technological route to synthesize the π phase shift. We use the dispersion of form birefringence of subwavelength gratings (SWGs), which are particularly adapted to longer wavelengths.

In this paper, we present the results of our work on the Annular Groove Phase Mask (AGPM) coronagraph, an achromatic VVC proposed in 2005 by our team (Mawet et al. 2005[1]). After several years of research, optimization and microfabrication - we're not at our first attempt - we managed to manufacture a promising component, dedicated to be integrated to the VISIR instrument at the VLT (June 2012). After a brief introduction to SWGs (Sect. 2), we show in Sect. 3 the design optimization of the VISIR-AGPM made out of diamond, with respect to manufacturability and performance. In Sect. 4 we briefly describe the microfabrication of the component, and we calculate the theoretical performances, which we expect to obtain on VISIR, using computer simulations based on the Rigorous Coupled Wave Analysis (RCWA). Finally, we conclude with the perspectives for present and future instruments.

## 2. THEORETICAL BACKGROUND

### 2.1 Subwavelength grating

The AGPM coronagraph is a micro-optical element that consists of a concentric circular subwavelength grating (SWG), thus a micro-optical structure with a period $\Lambda$ smaller than $\lambda/n$, $\lambda$ being the observed wavelength of the incident light and $n$ the refractive index of the grating substrate. In such a structure, only zeroth transmitted and reflected orders are allowed to propagate outside the grating. One can employ these SWGs to synthesize artificial birefringent achromatic wave plates (Kikuta et al. 1997[15], Nordin et al. 1999[16]) with two different refractive indices, $n_{TE}$ and $n_{TM}$, with regard to the polarization states TE (transverse electric) and TM (transverse magnetic), parallel and orthogonal to the grating grooves, respectively. The phase retardation $\Delta\Phi$ introduced by a birefringent SWG between the two polarization components is dependent on the wavelength, and is given by

$$\Delta\Phi_{TE-TM}(\lambda) = \left(\frac{2\pi}{\lambda}\right) h \, \Delta n_{form}(\lambda) \qquad (1)$$

where

$$\Delta n_{form}(\lambda) = n_{TE}(\lambda) - n_{TM}(\lambda) . \qquad (2)$$

$h$ is the optical path through the birefringent medium. In order to produce an achromatic wave plate, the product of the two factors in $\lambda$ on the right hand side of Eq. 2 needs to be a constant over a wavelength range as large as possible. By varying the grating parameters (geometry, material, incidence) the wavelength dependence in $\Delta n_{form}$ should be tuned to be closely proportional to the wavelength across a wide spectral band, in order to ensure the desired constant phase shift, in our case π.

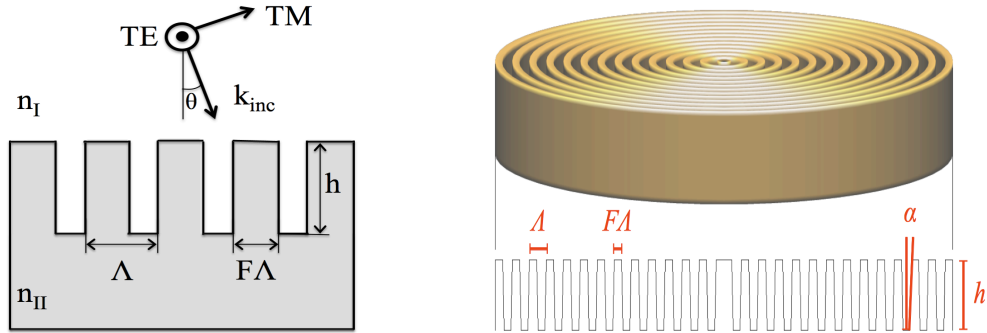

**Figure 1. Left**: Schematic diagram of a SWG. The incident light beam vector $k_{inc}$ is perpendicular to the grating lines, $\theta$ is the incidence angle and $n_I$ and $n_{II}$ are the refractive indices of the incident (superstrate) and transmitting (substrate) media, respectively. **Right**: The AGPM parameters definition. The geometry of the grooves is defined by their depth $h$, their period $\Lambda$ and their filling factor $F$, which corresponds to the ratio between the width of the lines and the period. The filling factor $F$ is such that $F\Lambda$ corresponds to the width of the grating walls.

The profile of the grooves is theoretically supposed to be rectangular. However, the performances of the microfabrication techniques required to fabricate such gratings are limited and the sides of the etched grooves are not perfectly vertical. Thus, another parameter to take into account in order to characterize the geometry of the grating is the slope $\alpha$ of the walls.

## 2.2 AGPM coronagraph working principle

The principle of the AGPM is illustrated in Fig. 2. It nominally works with unpolarized natural light. The collimated beam coming from the circular entrance pupil of the telescope converges in the focal plane where the mask is introduced and well-centered, so that the Airy disk (i.e. the Fourier transform of the pupil) is focused on the very center of the circular grating. In the focal plane, the mask affects the phase of the beam by a Pancharatnam $4\pi$ phase ramp. In fact, the AGPM consists of a space-variant subwavelength grating which synthesizes a vectorial optical vortex (vectorial means that the phase shift occurs between vectorial components $s$ and $p$) also called a Vector Vortex Coronagraph (VVC). Such a phase ramp corresponds to a topological charge $l = 2$ which analytically leads to a total starlight rejection in the theoretically perfect case (Foo et al. 2005[14], Mawet et al. 2005[1]). Behind the phase mask, the beam is collimated again inducing another Fourier transform of the electric field. In the next pupil plane, we can notice the effect of the mask which involves a perfect annular symmetry of the rejection around the original pupil. The Lyot stop, slightly undersized compared to the entrance pupil dimension, suppresses the diffracted starlight keeping only the central dark part. Lastly, the light is focused on a detector to produce the final coronagraphic image.

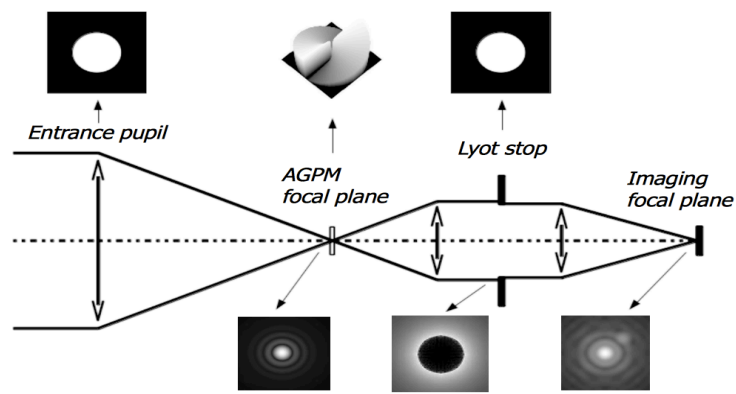

**Figure 2.** AGPM coronagraphic optical bench scheme. Numerical coronagraphic simulation illustrating the diffractive behavior of the AGPM in the *K*-band (Mawet et al. 2005[1]).

## 3. DESIGN OF AN AGPM FOR VISIR

Being part of the VISIR Upgrade Project (see Lagage *et al.* 2010[18] and Kerber *et al.* 2012[19], these proceedings) foreseen for June-August 2012, we have focused on the long-wavelength part of the N-band (11-13.2 μm). In fact, compared to bands of shorter wavelengths, the larger periods and lower aspect ratios of the N-band AGPM make this grating easier to fabricate (Delacroix *et al.* 2010[20]).

### 3.1 Material choice: diamond

Among the materials commonly used in the N-band, one good candidate is diamond, with a refractive index ~2.38 from 3 to 13 μm, covering the region of interest. The use of diamond substrates leads to many advantages. First and foremost, it has got a wide transmission window (Bundy 1962[21]). In addition, its mechanical properties are outstanding (low density = 3.52 kg/m$^3$ ; very high hardness = 10 on the Mohs scale ; very high elasticity). Also, its thermal (excellent conductor, inertia) and chemical (resistant to usual chemicals, acids, and most of the alkalies) properties make it an excellent candidate to be space qualified, which is of great interest for future missions. Some exotic materials with lower refractive index exist, but they are not compatible with our applications because of many disadvantages (brittle, deform easily, etc.). Materials with very high refractive index such as silicon (~3.4) are not compatible either, because they necessarily require complex antireflective coatings on top of the subwavelength grating.

### 3.2 Design optimization

The design of the grating was conducted in synergy with the manufacturing. In particular, the slope of the grating sidewalls must be taken into account and the aspect ratios must be kept within the range of what can be etched in the material. The etch process gives a slope $\alpha$ of ~2.7°. In the fabrication process, small errors in line width, slope and grating depth occur. In particular, the depth h and the slope $\alpha$ are difficult to measure precisely. A design which performs well even under small changes in these parameters was therefore sought. We have computed 2D maps of the Mean Null Depth over the upper N-band (11-13.2 μm) as a function of the filling factor ($F$) and of the depth ($h$), for several values of $\alpha$ ranging from 2.6 to 2.8°. We also calculated (Delacroix *et al.* 2012[22]) the mean and standard deviation of all these maps (see Fig. 3).

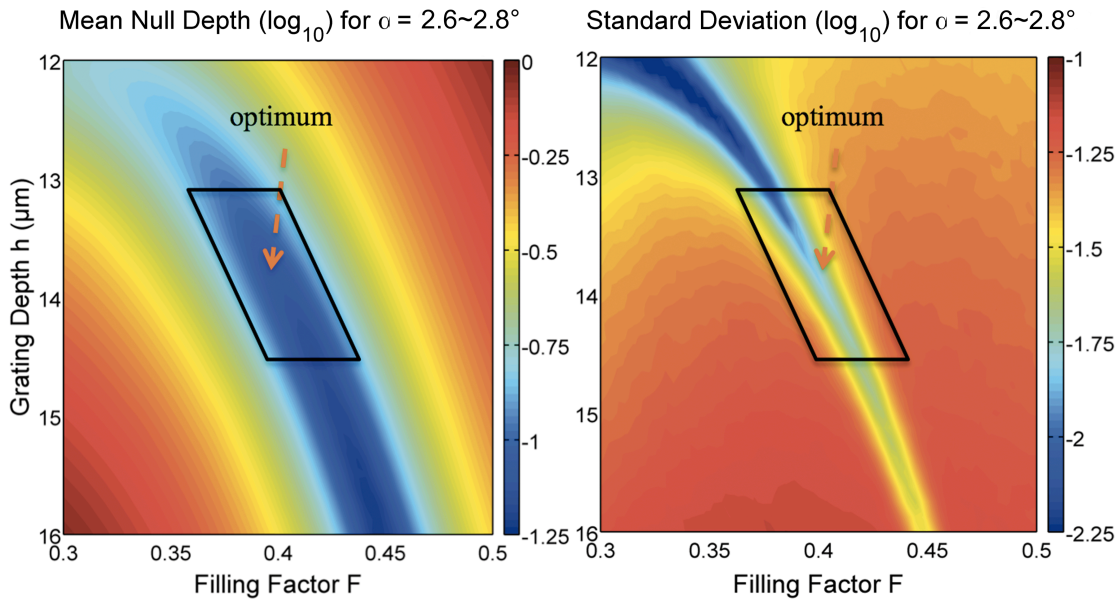

**Figure 3.** RCWA multi-parametric simulation: mean (left) and standard deviation (right) of the Mean Null Depth (logarithmic scale) over the upper N-band, with $\alpha$ ranging from 2.6 to 2.8°. The period is set to $\Lambda$ = 4.6 μm (SWG limit).

In order to simulate the grating response and to calculate its form birefringence $\Delta n_{TE-TM}$, and hence the Null Depth, we must consider the vectorial nature of light. Therefore, our numerical simulations have been performed using the Rigorous Coupled Wave Analysis (RCWA, Moharam & Gaylord 1981[23]), which resolves the Maxwell equations and gives the realistic entire diffractive characteristics of the simulated structure. To obtain a good compromise between the mean value and the standard deviation, the optimum values are $h = 13.86$ μm and $F = 0.4$ which correspond to a line width (on the top) $F\Lambda = 1.84$ μm.

## 4. MANUFACTURING AND PERFORMANCE ANALYSIS

### 4.1 Etching subwavelength and antireflective gratings

The manufacturing of the diamond VISIR-AGPM involves a concerted action with both the University of Liège (Belgium) and the Ångström Laboratory (Uppsala, Sweden). Based on our results of the design optimization study, we elaborated the technology stack and the optimal "recipe" for the microfabrication in order to achieve the required specifications. The fabrication method, based on the Swedish laboratory expertise (see Karlsson & Nikolajeff 2003[24], Karlsson *et al.* 2010[25]), involves nano-imprint lithography and reactive ion etching. The diamond substrate has been etched on both sides: one side with the circular subwavelength grating (SWG) which creates the Vector Vortex, and the other side with an antireflective grating (ARG) which maximizes the transmission (see Fig. 4).

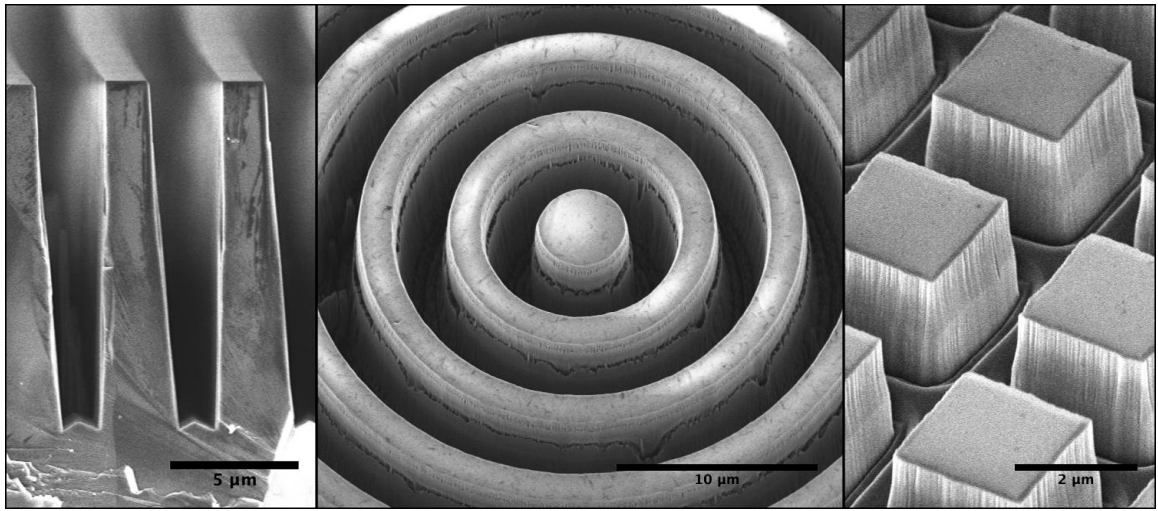

**Figure 4.** SEM-micrographs of a diamond AGPM for VISIR. **Left**: Cross sectional view of the grooves. **Middle**: centre of the AGPM. **Right**: Antireflective structure on the backside.

Incoherent reflections with different phase shift may interfere with the main beam and degrade performance of the component. The natural reflection of the diamond at N band is ~17% for one interface. The ARG was designed to reduce surface reflections using a programme based on RCWA to calculate values of the zero-order transmission. As can be seen on Fig 5., the transmittance of the final AGPM is in good agreement with the theoretical expectations. The total transmittance of the finished AGPM components ranges between 89 and 95% over the band.

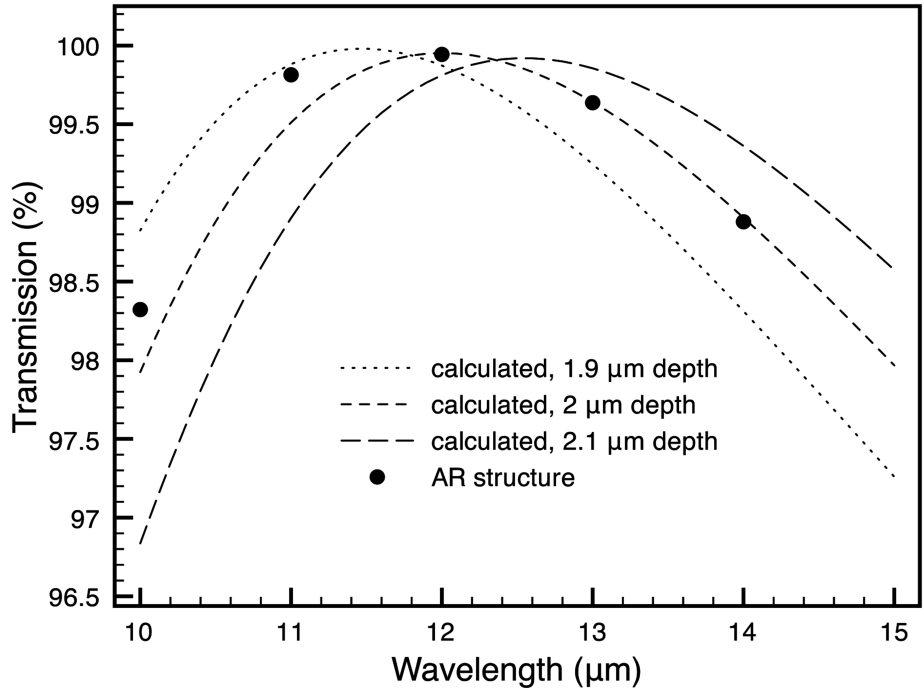

**Figure 5.** Transmission spectrum of one diamond interface with ARG measured with a spectrophotometer. The calculated transmission values for three different depths are also shown.

### 4.2 Centering marks

Besides the incoherent reflections which need an antireflective surface on the backside in order to be avoided, one must also develop a system for centering the AGPM. Indeed, the phase mask is very sensitive to tip-tilt, and it should be centered with a precision of several microns. One solution to help identifying the center of the vortex, is to add some fiducials (i.e. centering marks) visible while observing a flat field. Afterwards, the centering is done in real time by looking at the image on the detector. For the centering marks, we use Aerosol Jet Printing (AJP) which is a particularly innovative technology for the selective deposition of materials at micron-scale. We have printed 100 microns wide lines as shown in Fig. 6 with silver ink, and defined sintering conditions for optimum grip. Interferometric profilometry after cryogenic test showed no deterioration of the printed lines.

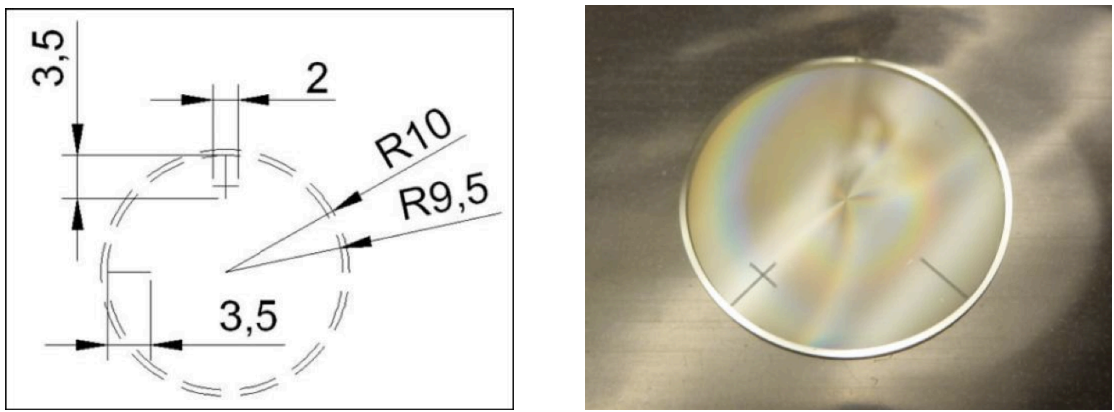

**Figure 6. Left**: Schematic diagram of the fiducials used for centering the AGPM. **Right**: Lines (centering marks) deposited by AJP on the VISIR-AGPM coronagraph.

### 4.3 Expected performances

Several AGPMs have been manufactured and a thorough metrology was conducted to select the component whose grating parameters are closest to the optimal specifications. An estimate of the depth was acquired by comparing several Scanning Electron Microscopy (SEM) images taken at different angles. The depth measured by this method on cracked samples was in good agreement with cross section images (within 2%). To reduce the aspect ratio of the grooves, we calculated the optimum with a fixed filling factor F=0.4 (see "Optim2" in Fig. 9). The so-called AGPM-N4 is the most efficient, with a Mean Null Depth ~$10^{-3}$ over the whole spectral band. This corresponds to a nulling ~$5.10^{-6}$ at *2λ/D* which is better than necessary for this specific application.

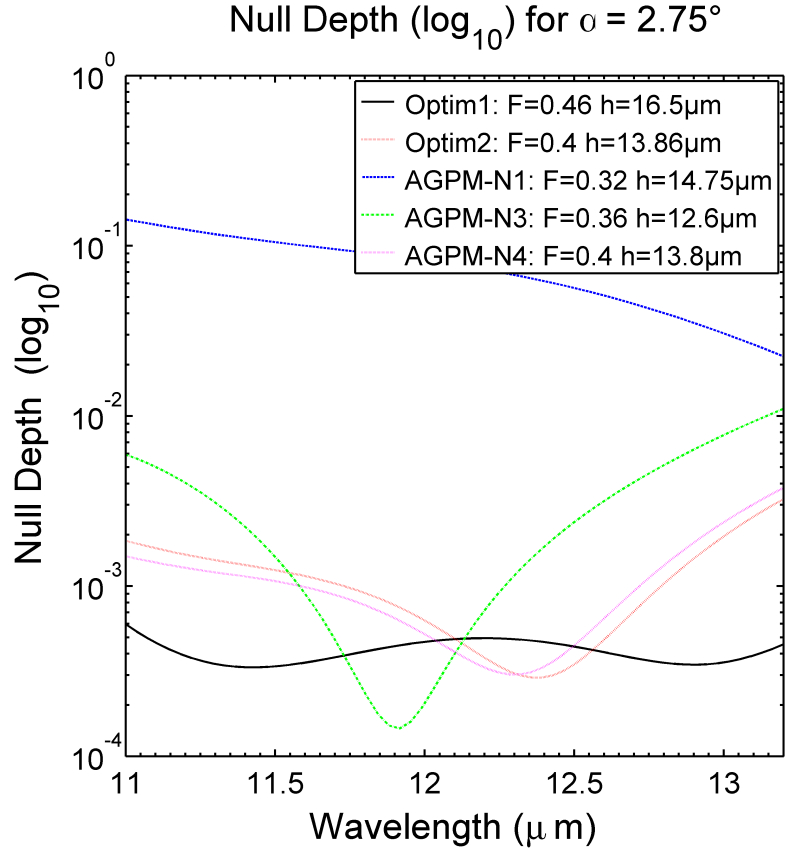

**Figure 9.** Calculated performances of several N-band AGPMs fabricated. Optimal specifications without constrain (Optim1) and with a constrained filling factor (Optim2).

## 5. CONCLUSION AND PERSPECTIVES

A diamond AGPM coronagraph was manufactured and is actually being integrated on the VISIR instrument at the VLT. It is provided with an antireflective grating etched on the backside of the component, allowing a total transmittance between 89 and 95% over the band. The centering of the device should be feasible thanks to fiducials printed with silver ink AJP. The VISIR-AGPM first light is scheduled for August 2012.

The diamond-optimized microfabrication techniques are more and more accurate despite the particularly high and unusual aspect ratio with regard to diamond microfabrication capabilities. Moreover, a new process using *e*-beam lithography is now being explored to reach smaller grating periods, and thereby shorter operating wavelengths. Future

AGPMs are being evaluated and developed for implementation on other high-contrast imaging instruments such as NACO (L-band ~3.8 µm) and SPHERE (K-band ~2.2 µm) at the VLT, or EPICS and METIS on the future E-ELT (Gilmozzi & Spyromilio 2007[26]).

# ACKNOWLEDGEMENT

C. Delacroix gratefully acknowledges the financial support of the Belgian Fonds pour la formation à la Recherche dans l'Industrie et dans l'Agriculture (FRIA) and Fonds de solidarité ULg. The authors from the Liège University acknowledge support from the Communauté française de Belgique - Actions de recherche concertée - Académie universitaire Wallonie-Europe.